# Adapting OC20-trained EquiformerV2 Models for High-Entropy Materials

Christian M. Clausen[a], Jan Rossmeisl[a], Zachary W. Ulissi[b,c]

[a]   Center for High-Entropy Alloy Catalysis (CHEAC), Department of Chemistry
      University of Copenhagen
      Universitetsparken 5, 2100 Copenhagen, Denmark
      E-mail: cmc@chem.ku.dk / jan.rossmeisl@chem.ku.dk
[b]   Department of Chemical Engineering
      Carnegie Mellon University
      5000 Forbes Avenue, Pittsburgh, PA 15213
[c]   Meta Fundamental AI Research
      250 Howard St, San Francisco, 94105, United States of America
      E-mail: zulissi@meta.com

**Abstract:** Computational high-throughput studies, especially in research on high-entropy materials and catalysts, are hampered by high-dimensional composition spaces and myriad structural microstates. They present bottlenecks to the conventional use of density functional theory calculations, and consequently, the use of machine-learned potentials is becoming increasingly prevalent in atomic structure simulations. In this communication, we show the results of adjusting and fine-tuning the pretrained EquiformerV2 model from the Open Catalyst Project to infer adsorption energies of *OH and *O on the out-of-domain high-entropy alloy Ag-Ir-Pd-Pt-Ru. By applying an energy filter based on the local environment of the binding site the zero-shot inference is markedly improved and through few-shot fine-tuning the model yields state-of-the-art accuracy. It is also found that EquiformerV2, assuming the role of general machine learning potential, is able to inform a smaller, more focused direct inference model. This knowledge distillation setup boosts performance on complex binding sites. Collectively, this shows that foundational knowledge learned from ordered intermetallic structures, can be extrapolated to the highly disordered structures of solid-solutions. With the vastly accelerated computational throughput of these models, hitherto infeasible research in the high-entropy material space is now readily accessible.

**Introduction**

The last decades of catalyst research have to a large extent focused on intermetallic surfaces with ordered crystal structures where only select sites are linked to their catalytic action. Quantum mechanical calculations, such as density functional theory (DFT), have been employed to investigate the underlying reaction energetics and subsequently leveraged to propose new catalyst materials[1–3]. However, this is only possible because ordered surfaces present a relatively finite number of unique binding sites for the adsorption of reaction intermediates. One of the main challenges when conducting computational research on high-entropy alloys (HEAs) and materials (HEMs) is, as the name suggests, the high degree of mixing entropy and the numerous microstates presenting an extremely heterogeneous surface with a staggering number of unique binding sites[4–6]. Rather than a handful of discrete adsorption energies, this results in distributions of adsorption energies determining the catalytic properties of the surface[7–9]. Obtaining a large enough sample of adsorption energies to represent these distributions can be prohibitively expensive, due to the computational cost of DFT. Furthermore, each distribution is tied to a specific material composition: As seen in Figure 1, due to the large number of constituent elements available to HEAs and HEMs, the number of possible catalyst materials rapidly grows beyond what can feasibly be screened computationally, let alone experimentally. If we are to truly explore the continuous compositional space that these materials offer, we need a significant speed-up of atomic structure simulations.

To this end, machine learning algorithms trained to emulate interatomic potentials (also known as MLPs) or directly infer material properties are becoming an increasingly relevant tool to enable high-throughput computations for bulk materials[10–12], biochemical systems[13], and catalyst surfaces[14–19]. Establishing and improving such algorithms has been the goal of the Open Catalyst Project since the publication of the OC20 dataset[20] and subsequent work[21–23]. The research team behind this effort outlined a series of goals for algorithms trained on OC20 to achieve, linking *Structure to Energies and Forces* (S2EF) and *Initial Structure to Relaxed Energy* (IS2RE), among others. The former, S2EF, substitutes a DFT single-point calculation of an adsorbate on a surface, yielding the forces on the atoms and the energy of the system. By iterative application, the relaxed atomic structure is obtained along with the adsorption energy. IS2RE, on the other hand, is the direct prediction of the adsorption energy from the initial atomic structure. This is significantly faster than traditional relaxation, but it is also a more difficult goal to achieve due to the multiple relaxation trajectories available to the adsorbate. S2EF and IS2RE are visualized in Figure 2a and 3a, respectively.

Here, we show that the EquiformerV2[24] (eqV2) algorithm trained on the OC20 dataset can yield state-of-the-art S2EF and IS2RE performance on out-of-domain HEA compositions with minor correction or fine-tuning. To this end, we have tested the inference performance of the algorithm on complex solid solution surfaces drawn from the composition space of Ag-Ir-Pd-Pt-Ru with adsorbed *OH and *O, which are relevant intermediates in the oxygen reduction reaction (ORR). We have focused on the mean absolute error (MAE) and mean error (ME) of adsorption energy prediction and have applied both the small and large pretrained versions of eqV2 with 31 and 153 million parameters, respectively (eqV2-31M and eqV2-153M).

**Methods**

The dataset of DFT simulated fcc(111) 3x3x5 atom-sized slabs of Ag-Ir-Pd-Pt-Ru used for fine-tuning and testing the models were carried out using ASE[25] and GPAW[26] software suites and have been published as part of previous work[27] where computational details can be found. Only calculations where *OH and *O are adsorbed to on-top or fcc-hollow sites, respectively, were included in the study. The DFT-calculated adsorption energies were split using an 80:10:10 ratio. The training set included 240 unique slabs with a total of 2031/1884 adsorptions of *OH/*O. The validation set included 30 unique slabs with a total of 253/230 adsorptions of *OH/*O. The test set included 31 unique slabs with a total of 260/237 adsorptions of *OH/*O.

To correct the model output of the S2EF models, the individual energy contributions of the surface atoms *not* directly coordinated to the adsorbate were subtracted from the total predicted adsorption energy. Coordination between atoms were determined based on the overlap of their covalent radii as seen in table S1.

Fine-tuning of the S2EF models utilized all structures in the relaxation trajectories of the DFT-calculated training and validation sets. Where an S2EF model was used to calculate samples that would subsequently be used to train an IS2RE model, the training and validation set always comprised 20,000 and 5,000 samples. Table S2 and S3 provides an overview of the number of samples in the respective sets. The energies of the S2EF samples are calculated as:

$$\Delta E^{DFT}_{*ads} = E^{DFT}_{slab+ads} - E^{DFT}_{slab} - E^{DFT}_{ads}$$

with $E^{DFT}_{slab+ads}$ being the total energy of the structure, $E^{DFT}_{slab}$ being the total energy of the fully relaxed slab, and $E^{DFT}_{ads}$ being the total energy of the gas phase reference which was $E^{DFT}_{H_2O} - \frac{1}{2}E^{DFT}_{H_2}$ for *OH and $E^{DFT}_{H_2O} - E^{DFT}_{H_2}$ for *O.

The IS2RE samples were constructed in such a way that all slabs have the exact same structure, but the element identities differ. This effectively removes all detailed positional information of the atoms except for the adsorbate, which can only be found in fixed locations (on-top or fcc-hollow) at a set height (2.0 and 1.3 Å). This represents completely identical initial structures (even identical cell dimensions) to use for IS2RE fine-tuning and testing.

Fine-tuning of the S2EF and IS2RE models was capped at 25 and 80-100 epochs, respectively, and the batch size was 2 for all models. Further details on the fine-tuning of models can be found in the available configuration files.

Comparing the inference speeds of the S2EF and IS2RE model was based on the entire dataset of HEA samples. Using DFT to relax a slab requires a mean compute of 87.97 CPU hours and assuming 8 CPU cores (16 threads) per job, which was used for this dataset, that equals 5.50 hours of walltime. This estimate is not including the extra time required to relax the slab without adsorbate as well. Using the pre-trained eqV2-153M S2EF model to relax the slabs (batch size of 2) to the same convergence criteria took a mean GPU walltime of 3.20 seconds and finally the mean inference speed (batch size of 8) of the fine-tuned eqV2-31M S2EF model was 0.024 seconds. The difference in batch sizes was due to the memory limitation of a single GeForce RTX 3090 GPU. The CPUs used were a mix of Intel Xeon E5-2650 v3/v4 and Intel Gold 6230R/6248R.

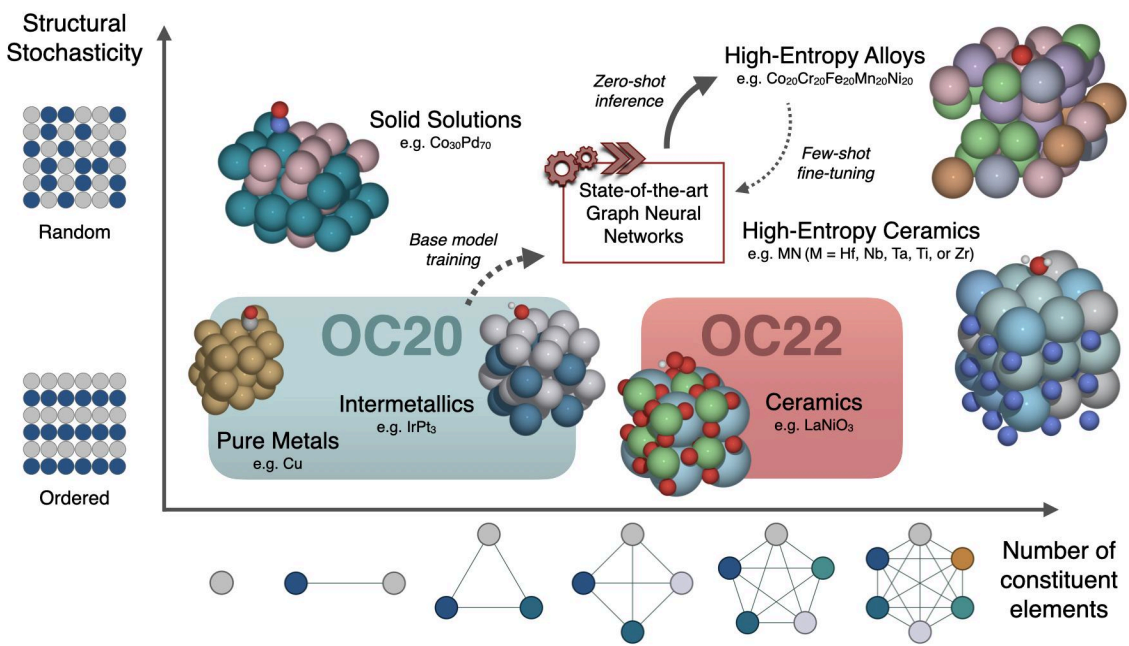

**Figure 1** Visualization of different catalyst material classes. The complexity and multitude of possible atomic surface environments increases sharply with disorder (shown on the y-axis) and number of elements (shown with orthographic projections on the x-axis). The OC20 dataset[20] is composed of ordered intermetallic structures drawn from the Materials Project[28] with up to three constituent elements. Here, we show that OC20-trained MLPs and adsorption energy inference algorithms, specifically the EquiformerV2 model[24], can extrapolate to the out-of-domain composition space of HEAs (*zero-shot inference*) with minor corrections and yield accurate inference on the ORR-relevant intermediates, *OH and *O. Alternatively, a set of DFT calculated structures of HEAs can be used to obtain even greater accuracy (*few-shot fine-tuning*).

**Results and Discussion**

Initially, we investigated the S2EF capabilities by predicting the adsorption energies of HEA samples already relaxed with DFT, equivalent to a single-point calculation. Figure 2b shows the result of the pretrained eqV2-153M model. As seen, there is a systematic offset towards weaker bonds for both the *OH and *O adsorption energies, as well as a spread of outliers resulting in subpar performance. To find the source of this offset, we turned our attention to the energy contribution from each node in the graph representation of the atomic environment. The eqV2 architecture ultimately sums these contributions to form the predicted adsorption energy. At closer inspection, we found that some atoms in the surface and subsurface layers had non-negligible contributions even though they were situated at a distance or angle from the binding site where little electronic perturbation of the binding site would be expected[29]. This is logical when considering that the OC20 dataset does not contain any HEMs but rather structured surfaces pulled from the Materials Project[28]. These systems consist of recurring patterns in both the horizontal and vertical directions, and therefore information about the binding site is delocalized throughout the cell. Contrarily, this is not the case with the stochastic structures of HEAs. To address this, we used a filter to only include the energy contributions from "*Atoms of Interest*" directly coordinated to the adsorbate molecule determined by cut-off radii, as illustrated in Figure 2c. This drastically improves the model output, as seen in Figure 2d, especially for *OH which displays an impressive MAE of 0.030 eV. The filter works to a lesser extend for adsorption energies of *O, that still has an ME of 0.060 meV. This is likely related to the three-fold fcc hollow binding site of *O. However, other filter schemes, such as only including the contribution of the nearest surface atom, worsened the performance on solid solution surfaces with fewer constituent elements, as seen in Figure S1. Thus, this appears to be the best possible compromise to encompass alloys with both high and low degrees of element complexity. Optionally, the eqV2-153M S2EF model can be fine-tuned to supersede the need for the filter by re-training the model on HEAs. We investigated this using increasing numbers of HEA samples to establish the learning curves seen in Figure S3. Here, we observe that even at small training set sizes, the S2EF performance of *O is markedly improved. For *OH, the prediction accuracy is actually slightly worsened unless enough HEA samples are provided, which testifies to the extensive size of the OC20 dataset.

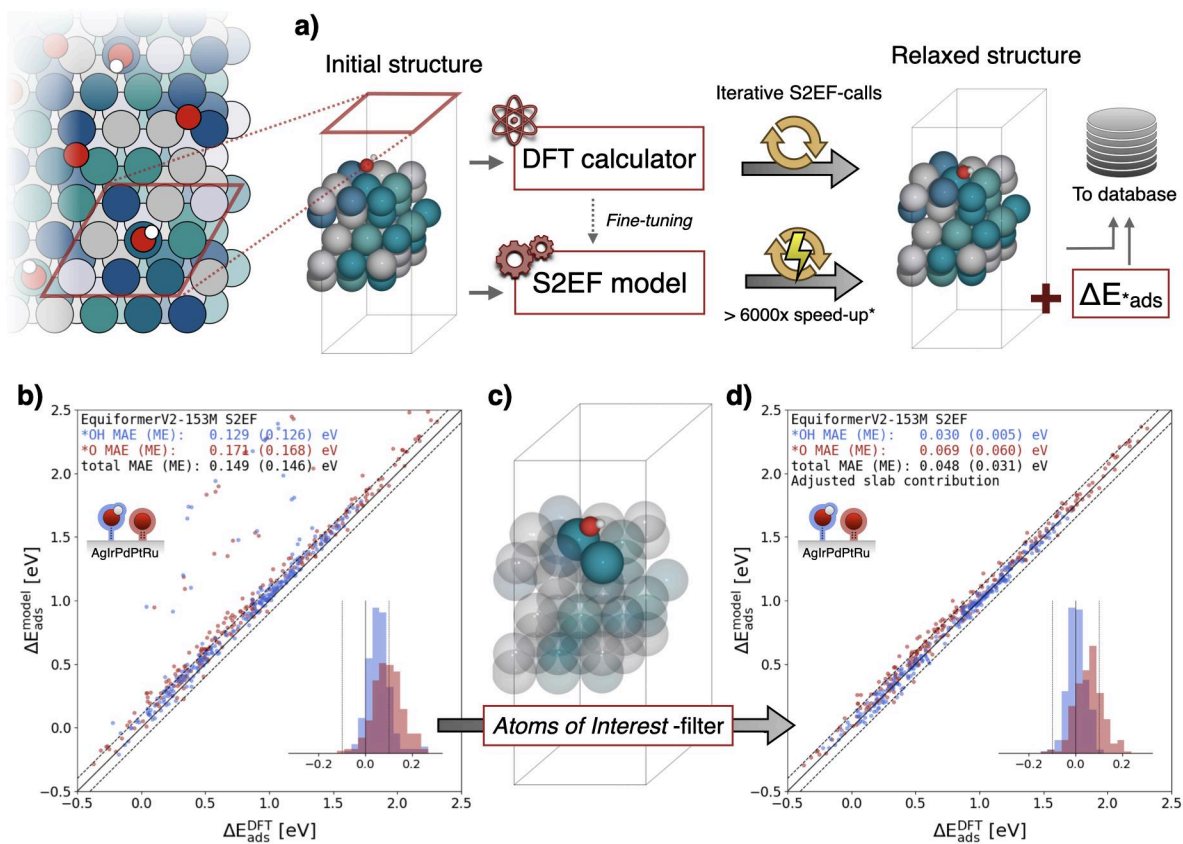

**Figure 2 a)** A schematic overview of atomic structure optimizations and adsorption energy ($\Delta E_{*ads}$). From an HEA surface, a unit cell is extracted. With a DFT calculator, the total energy of this structure and the forces acting on the atoms can be calculated. By doing this iteratively, the structure is optimized and the adsorption energy of the molecule is obtained. The DFT calculator can be replaced by the significantly faster machine-learned S2EF model, which yields the adsorption energy and forces. **b)** Parity plot of predicted adsorption energies for *OH (blue) and *O (red) using the pretrained eqV2-153M S2EF model. Dashed lines mark errors of ± 0.100 eV and the insert displays the distribution of errors. The energies stem from a single S2EF call on DFT-relaxed structures to allow for direct comparison. **c)** Visualization of the filter correcting the adsorption energies by solely including the energy contributions from surface atoms directly coordinated to the adsorbate molecule. Contributions from all other surface atoms are excluded (shown as transparent) **d)** Parity plot after application of the filter. *For details on the speed-up estimate see the Methods section.

As mentioned earlier, the alternative to iterative structure relaxation is an IS2RE model that solely provides the adsorption energy in a single call. These models are trained on the initial unrelaxed structures labeled with the adsorption energies of the corresponding relaxed structures, either obtained with DFT calculations or an S2EF model. The latter results in the hierarchical set-up of a general MLP informing a more specialized direct prediction model in a process akin to knowledge distillation[30]. To obtain such a model, we trained various different eqV2 models for IS2RE. As no pretrained eqV2 IS2RE model was available, the onset was the pretrained S2EF model, which was trained for the IS2RE task on a series of different training sets. These sets were obtained with either DFT (*DFT→IS2RE*) or with the eqV2-153M S2EF model (*S2EF→IS2RE*). This model could itself be fine-tuned with DFT data (*DFT→S2EF→IS2RE*), as illustrated by the flow chart in Figure 2b. The latter configuration may seem like unnecessary steps, but using *few-shot* fine-tuning (implying few DFT samples) to obtain an accurate S2EF model that can subsequently make a training set of tens of thousands relaxations can be beneficial, as our results in Figure 2c show: Not only do we see that the model is readily trained for the IS2RE with only a few hundred DFT samples but also performs admirably when trained on the filtered *zero-shot* output of the pretrained eqV2-153M S2EF model. Interestingly, using the layered approach (*DFT→S2EF→IS2RE*) results in visibly improved accuracy for *O but not for *OH, indicating that key features of the *OH on-top binding site are more easily captured than the more complex three-fold fcc hollow site of *O. Therefore, the large training set created with the fine-tuned S2EF model grants a boost in accuracy as more permutations of the hollow sites are explored. As seen in Figure S4, we observe little difference between the

eqV2-31M and eqV2-153M when choosing a model to train for the IS2RE task. Thus, the smaller model with fewer parameters is adequate for this application and is our preferred choice based on the lower memory requirements and resulting larger batch size during inference.

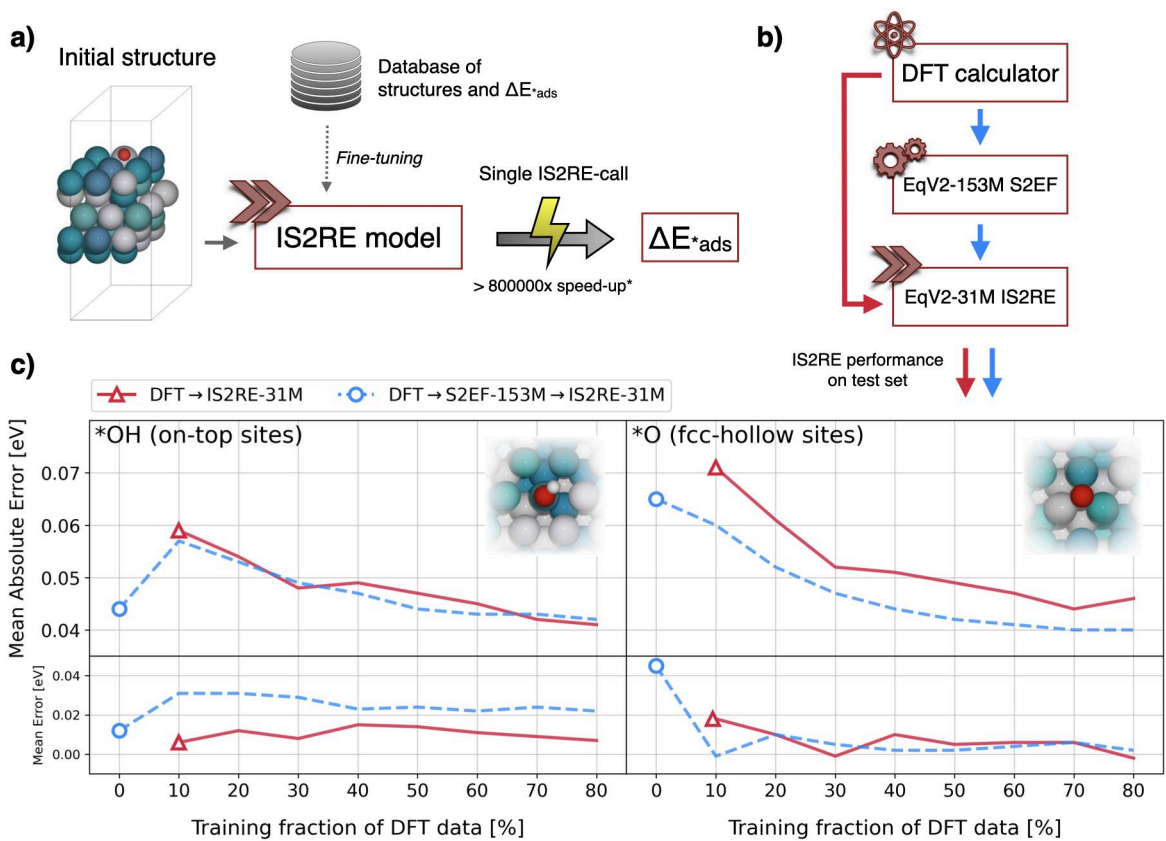

**Figure 3 a)** Using structures and adsorption energies from either a DFT calculator or S2EF model, an IS2RE model can be fine-tuned to directly map initial structure to adsorption energy, enabling extremely high-throughput applications. **b)** Schematic showing how DFT and S2EF output was used to fine-tune the different IS2RE models. The *DFT→IS2RE* models were trained directly on DFT data. The *DFT→S2EF→IS2RE models* were trained on the output of the eqV2-153M S2EF model fine-tuned on DFT data. However, the "0%"-model used the output of pretrained eqV2-153M S2EF model with the "*Atoms of Interest*" energy correction shown in Figure 2c. **c)** Prediction accuracy of different IS2RE models for *OH and *O adsorbed in on-top and fcc-hollow sites, respectively, with MAE plotted in the upper panels and ME plotted in the lower panels. The x-axes note the size of the training set used to fine-tune the model.

In Figure S4, we also show the performance of an IS2RE model applied in some of our earlier publications[9,27,31]. This is a very lean graph neural network (GNN) with only 3062 parameters so inference speed is practically instantaneous. However, even though it attains a decent accuracy, especially when trained on enough data by including the S2EF model in the set-up (*DFT→S2EF→leanGNN*), the eqV2 IS2RE model still achieves ~30% lower MAEs. A hyperparameter optimization of the lean GNN could improve the performance and might be worth pursuing in future work for extreme high-throughput applications.

**Conclusion**
To conclude, we have shown that the OC20-trained eqV2-153M S2EF model is able to perform zero-shot inference on out-of-domain solid solutions and HEA systems with state-of-the-art accuracy by filtering the adsorption energy output based on the relevant atoms constituting the binding site. Alternatively, fine-tuning the S2EF model renders the filtering obsolete and already leads to low inference errors with little training data. The same is true when fine-tuning the models for IS2RE tasks yielding errors as low as ~0.040 eVs for the investigated system. Additionally, the energetic and structural output of the S2EF models are accurate enough to serve as a foundation to train IS2RE models in a knowledge distillation set-up. In combination with S2EF fine-tuning, this is observed to boost the accuracy on *O in fcc-hollow sites compared to direct IS2RE fine-tuning.

Used as a replacement for a DFT calculator, these pretrained models potentially cut data acquisition time from months to days or even hours. In the early stages of a high-throughput research project these can be employed as probing tools before committing significant computational resources to DFT calculations. Fine-tuned to the specific domain-of-interest, they can deliver accurate adsorption energy inference at a rate that would have required an unthinkable amount of computing resources. Both approaches will be valuable in facilitating computational screening of catalyst materials in high-dimensional composition space and will enable research otherwise impossible due to the degree of disorder in HEAs and HEMs.

**Data availability**
The OCP framework and the pre-trained weights for the models are available at the OCP github repository. Parity plots of all reported results, as well as config-files, data processing, DFT calculations and more will be available upon manuscript acceptance at a publicly available link.

**Acknowledgements**
The authors extend their gratitude to Abhishek Das, Brandon Wood, Muhammed Shuaibi, and Larry Zitnick for their insightful comments and feedback. CMC and JR acknowledge support from the Danish National Research Foundation Center for High-Entropy Alloy Catalysis (CHEAC) DNRF-149.

**Table of Contents Image**

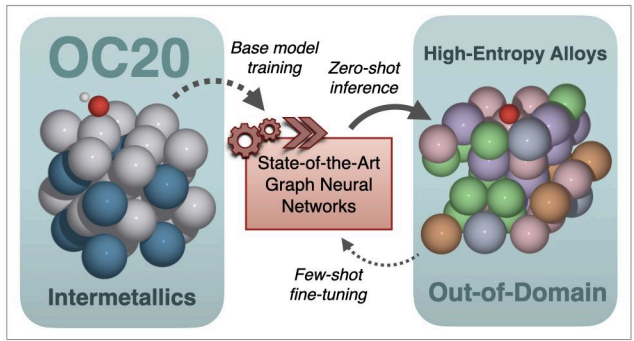

# Supplementary information

## Adapting OC20-trained EquiformerV2 Models for High-Entropy Materials

Christian M. Clausen[a], Jan Rossmeisl[a], Zachary W. Ulissi[b,c]

[a]  Center for High-Entropy Alloy Catalysis (CHEAC), Department of Chemistry
     University of Copenhagen
     Universitetsparken 5, 2100 Copenhagen, Denmark
     E-mail: cmc@chem.ku.dk / jan.rossmeisl@chem.ku.dk
[b]  Department of Chemical Engineering
     Carnegie Mellon University
     5000 Forbes Avenue, Pittsburgh, PA 15213
[c]  Meta Fundamental AI Research
     250 Howard St, San Francisco, 94105, United States of America
     E-mail: zulissi@meta.com

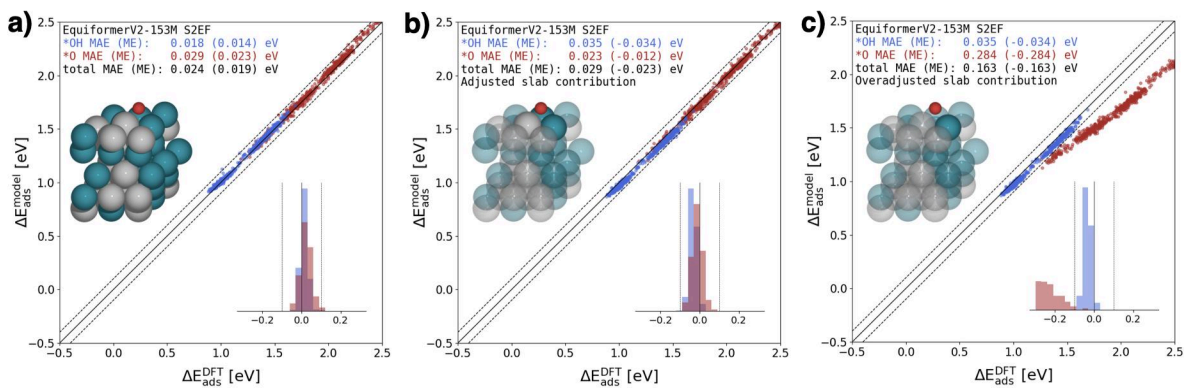

**Figure S1** Parity plots of predicted adsorption energies for *OH (blue) and *O (red) on a separate test set of Ag-Pd surfaces using the pretrained eqV2-153M S2EF model. The mean absolute error (MAE) is displayed, with the mean error (ME) following in parenthesis. Dashed lines mark errors of ± 0.100 eV and the insert displays the distribution of errors. **a)** Unadjusted output of the model. **b)** The output after applying the "*Atoms of Interest*"-filter subtracting the contribution of all atoms not coordinated to the adsorbate molecule from the predicted adsorption energy. **c)** The overadjusted output, only including the contribution of the surface atom closest to the adsorbate. Inserts show a visualized slab with the opaque atoms contributing to the model output and the transparent atoms not contributing.

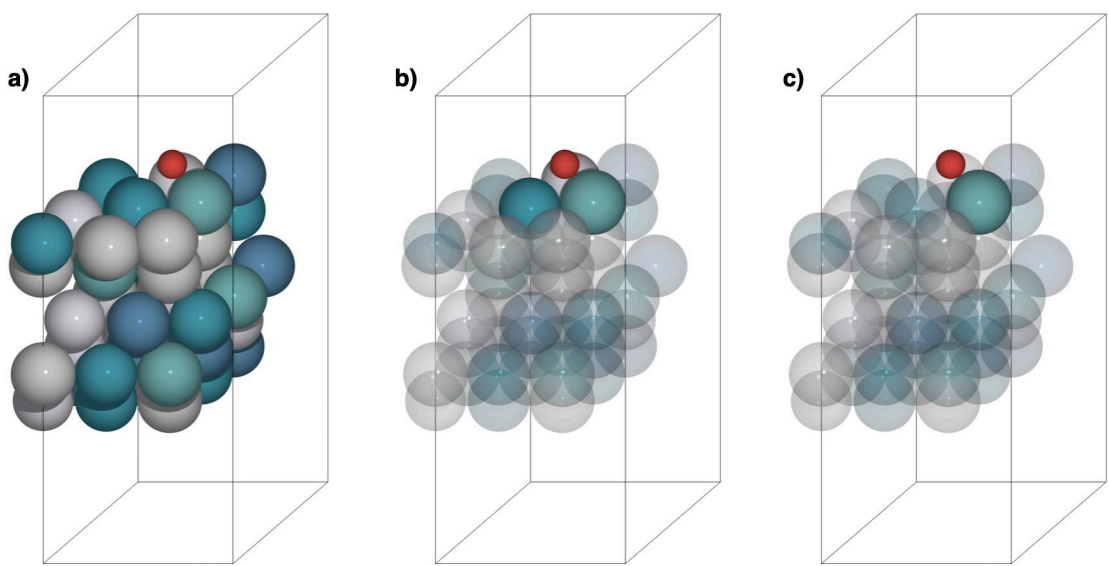

**Figure S2** The "*Atoms of Interest*"-filter visualized on a sample slab of the Ag-Ir-Pd-Pt-Ru solid solution surface. **a)** The energy contribution of all atoms in the cell is included. This is the default output of the eqV2 model. **b)** Only the energy contribution of the atoms directly coordinated to the adsorbed *O is included along with the contribution of O itself. These atoms are illustrated as opaque, while the atoms not included are translucent. **c)** Only the energy contribution of the adsorbate and the closest surface atom is included, resulting in an overcorrected adsorption energy as seen in Figure S1c.

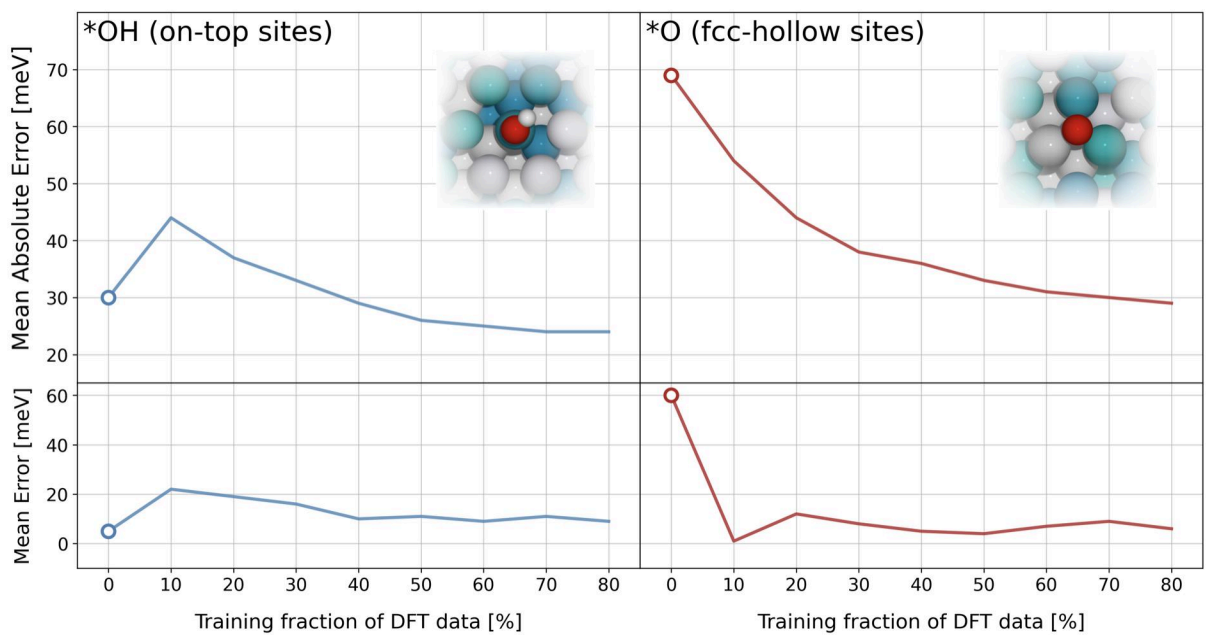

**Figure S3** Prediction accuracy of the eqV2-153M S2EF model for *OH and *O adsorbed in on-top and fcc-hollow sites, respectively, with MAE plotted in the upper panels and ME plotted in the lower panels. The "*Atoms of Interest*" energy correction shown in Figure 2c was applied when not fine-tuned on DFT data (points corresponding to 0% on the x-axes). Otherwise the S2EF model was fine-tuned on the fraction of the DFT dataset as listed on the x-axes.

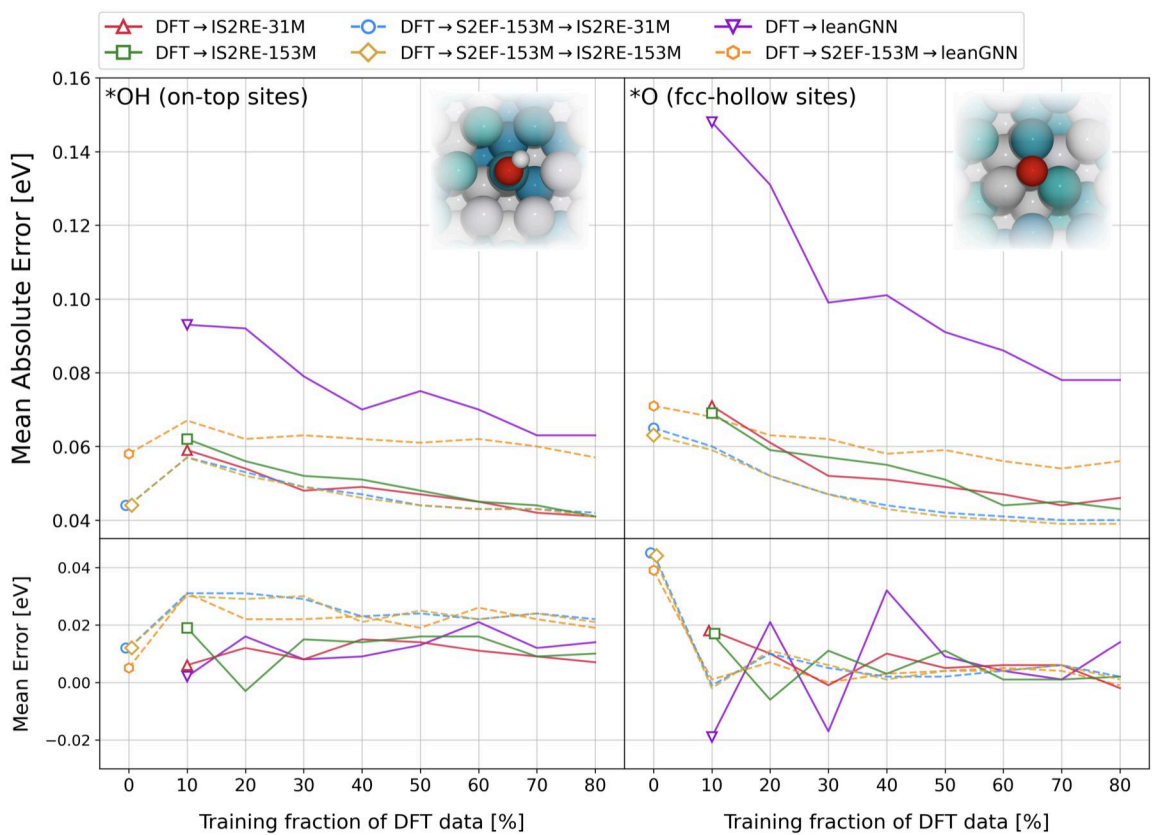

**Figure S4** Prediction accuracy of different IS2RE models for *OH and *O adsorbed in on-top and fcc-hollow sites, respectively, with MAE plotted in the upper panels and ME plotted in the lower panels. The x-axes note the size of the training set used to fine-tune the model. An overview of what data was used to train/fine-tune the model are shown in Figure 3b.

**Table S1** Covalent radii of elements for determining coordination in atomic structures.

| Element | Radii [Å] |
|---------|-----------|
| Ag | 1.45 |
| Ir | 1.41 |
| Pd | 1.39 |
| Pt | 1.36 |
| Ru | 1.46 |
| O  | 0.66 |
| H  | 0.31 |

**Table S2** Overview of DFT-calculated samples in the training, validation, and test sets. The slabs represent a 3x3 atom-sized surface hence they can yield up to nine different on-top and fcc-hollow adsorption sites.

| Fraction of DFT data | Number of unique slabs | *OH / *O IS2RE samples | *OH / *O S2EF samples |
|----------------------|------------------------|------------------------|------------------------|
| 10% | 30 | 253 / 235 | 3052 / 1610 |
| 20% | 60 | 512 / 477 | 6031 / 3220 |
| 30% | 90 | 768 / 716 | 8906 / 4854 |
| 40% | 120 | 1021 / 956 | 11862 / 6540 |
| 50% | 150 | 1267 / 1177 | 14968 / 8132 |
| 60% | 180 | 1521 / 1408 | 18161 / 9793 |
| 70% | 210 | 1769 / 1624 | 21103 / 11341 |
| 80% | 240 | 2031 / 1884 | 24056 / 13057 |
| Validation (10%) | 30 | 252 / 230 | 2906 / 1597 |
| Test (10%) | 31 | 260 / 237 | - |

**Table S3** Overview of S2EF calculated samples in the training and validation sets.

| S2EF model | *OH / *O IS2RE training samples | *OH / *O IS2RE validation samples |
| --- | --- | --- |
| S2EF-31M | 9769 / 10231 | 2434 / 2566 |
| S2EF-153M | 9527 / 10473 | 2402 / 2598 |
| $DFT_{10\%} \to$ S2EF-153M | 9969 / 10031 | 2471 / 2529 |
| $DFT_{20\%} \to$ S2EF-153M | 9878 / 10122 | 2461 / 2539 |
| $DFT_{30\%} \to$ S2EF-153M | 9972 / 10028 | 2477 / 2523 |
| $DFT_{40\%} \to$ S2EF-153M | 9924 / 10076 | 2470 / 2530 |
| $DFT_{50\%} \to$ S2EF-153M | 9776 / 10224 | 2411 / 2589 |
| $DFT_{60\%} \to$ S2EF-153M | 9532 / 10468 | 2360 / 2640 |
| $DFT_{70\%} \to$ S2EF-153M | 9473 / 10527 | 2339 / 2661 |
| $DFT_{80\%} \to$ S2EF-153M | 9470 / 10530 | 2351 / 2649 |